\documentclass[a4paper,11pt]{article}
\pdfoutput=1 

\usepackage{jinstpub} 

\usepackage{liecolon}

\usepackage{subcaption}

\title{\boldmath Averaged Invariants in Storage Rings with Synchrotron Motion}

\author[a,1]{S. Webb,\note{Corresponding author.}}
\author[a]{N. Cook,}
\author[b]{J. Eldred}

\affiliation[a]{RadiaSoft LLC,\\6525 Gunpark Dr., Suites 370-411, Boulder, CO, USA}
\affiliation[b]{Fermi National Accelerator Laborator,\\Batavia, IL, USA}

\emailAdd{swebb@radiasoft.net}

\abstract{In an ideal accelerator, the single-particle dynamics can be decoupled into transverse motion -- the betatron oscillations -- and longitudinal motion -- the synchrotron oscillations. Chromatic and dispersive effects introduce a coupling between these dynamics, the so-called synchro-betatron coupling. We present an analysis of the fully coupled dynamics over a single synchrotron oscillation that leads to an averaged invariant with synchro-betatron coupling in a generic lattice. We apply this analysis to two problems: first, a toy lattice where the computations are analytically tractable, then a design for a rapid cycling synchrotron built using the integrable optics described by Danilov and Nagaitsev, showing that although there is fairly complex behavior over the course of a synchrotron oscillation, the Danilov-Nagaitsev invariants are nevertheless periodic with the synchrotron motion.}


\arxivnumber{2007.04935} 

\begin{document}
\maketitle
\flushbottom

\section{Introduction}
\label{sec:intro}

Single-particle dynamics in particle accelerators can be broken into the fast transverse betatron oscillations, with tunes $\nu_{x, y} \gg 1$, and the much slower synchrotron oscillations, with tunes $\nu_s \ll 1$. For a coasting beam, the momentum dependence of the focusing element strengths leads to chromaticity, a momentum-dependent betatron tune, and dispersion, a momentum-dependent closed orbit. When an rf cavity is added and synchrotron motion occurs, that synchrotron motion couples to the betatron motion through the chromaticity and dispersion -- so-called synchro-betatron coupling.

Synchro-betatron coupling can lead to complex coupled dynamics. The impact of synchro-betatron coupling has been well-studied for linear alternating gradient focusing lattices~\cite{hammer:1955,orlov:1957,lee:94}, but their influence on more novel lattice designs, such as nonlinear integrable optics~\cite{danilov_nagaitsev:10,nag_val_dan_10,nag_dan_ipac:10,nagaitsev_valishev:10}, has yet to be studied in detail.

In this paper, we calculate a stroboscopic invariant of coupled synchro-betatron motion in the limit of small synchrotron tune. This invariant is the Hamiltonian that generates an $N$-turn map, where $N \nu_s \approx 1$. We show that this Hamiltonian is correct to $\mathcal{O}(\nu_s)$, and that the perturbing terms do not cause secular growth in the invariants. This Hamiltonian is a pure function of the transverse coordinates and the synchrotron action coordinate -- thus if this Hamiltonian is integrable then the entire system is integrable over $N$ turns. We demonstrate the preservation of integrable dynamics in the context of an integrable rapid cycling synchrotron, designed to use nonlinear integrable optics to mitigate beam loss due to coherent instabilities in high intensity proton beams. This result relies on the single-turn Lie map formalism of Dragt \emph{et al.}~\cite{dragt_abell:96,dragt_finn:1976,dragt_forest:83,dragt_text,dragt:79,dragt:82,dragt:84}, and therefore we give a brief survey of key results in Appendix~\ref{app:sym_maps}.

\section{Synchro-betatron Coupling}
\label{sec:maps}

The simplest model for single-particle dynamics in a particle accelerator is the uncoupled vertical and horizontal betatron oscillations, with independent synchrotron motion longitudinally. Dispersion can complicate this picture, as each trajectory's momentum-dependent closed orbit oscillates with the synchrotron motion. The simplest linear cases create normal modes that couple transverse and longitudinal motion, and the eigenemittances and tunes can computed with a generic symplectic matrix formulation~\cite{forest:98}. The next-leading-order dynamics result from chromaticity, the momentum dependence of the betatron tune. Because the system is Hamiltonian, a betatron tune that depends on the momentum implies a betatron amplitude dependence in the synchrotron tune -- so-called \emph{synchro-betatron coupling}.

This coupling has a number of implications. The slow change in the momentum offset of a particle's trajectory suggests that there will also be a slow change in the chromaticity -- the synchrotron motion will modulate the betatron oscillations with a frequency of the betatron phase. Synchro-betatron coupling therefore can lead to sidebands in the betatron motion located at $\nu_\perp \pm m \nu_s$ for betatron tune $\nu_\perp$ and synchrotron tune $\nu_s$~\cite{hammer:1955,orlov:1957}. Because of the coupling in the system, the synchrotron motion modifies the usual transverse action-angle variables~\cite{lee:94}. This can lead to synchro-betatron coupling induced parametric resonances when the betatron tune is a harmonic of the synchrotron tune.

Synchro-betatron coupling is conceptually similar to adiabatic analysis in that a quantity that affects the transverse motion is changing slowly compared to the transverse oscillations. Because the synchrotron oscillations are slow and periodic, we expect to be able to find a period-averaged Hamiltonian treatment of synchro-betatron coupling. However, because the slowly changing quantity is a dynamical quantity in a Hamiltonian system, the analysis is more subtle -- we must make sure any treatment of synchro-betatron coupling reflects the Hamiltonian nature of the dynamics.

\section{A Stroboscopic Hamiltonian}
\label{sec:strobo}

Synchrotron motion in the absence of coupling is a periodic system, slowly varying compared to the much faster betatron oscillations. The periodicity of the synchrotron motion suggests examining the total dynamics stroboscopically, looking every $N$ turns where $N \times \nu_s \approx 1$ to analyze the synchrotron-period-averaged influence on the transverse dynamics.

Suppose we have a storage ring comprised of a sequence of transverse elements and a single, thin rf cavity. The single-turn map takes the form
\begin{equation}
\mathcal{M} = \mathcal{M}_{\delta} \mathcal{M}_{V}
\end{equation}
where the Hamiltonian that generates $\mathcal{M}_{\delta}$ is of the form
\begin{equation}
H_{\delta} = H_\perp(\vec{z}_\perp, \delta) + \alpha_c(\delta)
\end{equation}
and the rf potential that generates the thin cavity map $\mathcal{M}_V$ is generated by $V(\phi)$. Here $\delta$ and $\phi$ are canonically conjugate, $\alpha_c$ captures the momentum compaction of the ring, and $H_\perp$ describes the transverse motion with chromatic and dispersive effects. Because $\alpha_c$ commutes with $H_\perp$, we can factor this into three maps: $\mathcal{M}_\perp \mathcal{M}_c \mathcal{M}_{V}$, where $H_\perp$ generates $\mathcal{M}_\perp$, $\alpha_c$ generates $\mathcal{M}_c$, and $\mathcal{M}_V$ is as before. In the absence of synchro-betatron coupling, the transverse dynamics are specified entirely by $\mathcal{M}_\perp$ and the longitudinal dynamics are specified entirely by the synchrotron map $\mathcal{M}_s = \mathcal{M}_c \mathcal{M}_{V}$.

We assume the synchrotron motion is integrable, so the dynamics can be specified in action-angle coordinates $(A, \psi)$, and 
\begin{equation}
\mathcal{M}_s = \exp \{ - \lieop{h_s(A)} \}
\end{equation}
with amplitude-dependent synchrotron phase advance $\mu_s(A)$. Because the synchrotron motion is presumed integrable, we know that $\delta$ must be a periodic function with the synchrotron phase, and can be written as a Fourier series
\begin{equation}
\delta = \sum_m \delta_m(A)~ e^{i m \psi}.
\end{equation}
The full single-turn map is therefore
\begin{equation}
\mathcal{M} = \exp \left \{ - \lieop{H_\perp(\vec{z}_\perp; A, \psi)} \right \} \exp \{ - \lieop{h_s(A_s)} \}.
\end{equation}
Once again, because $\delta$ is periodic with the synchrotron phase, so too is $H_\perp$, and we can rewrite
\begin{equation}
H_\perp = \sum_{k=-\infty}^{\infty} h_k(\vec{z}_\perp, A)~e^{i k \psi} = \overline{H}(\vec{z}_\perp, A) + \sum_{k \neq 0} h_k(\vec{z}_\perp, A)~e^{i k \psi}
\end{equation}
where we have called out $\overline{H} = \langle H_\perp \rangle_{\psi_s}$, the average of $H_\perp$ over the synchrotron phase, as it will be important later. That $H_\perp$ is a real function requires that $\overline{H}$ be real, and that $h_{-k} = h_k^{*}$.

For stroboscopic dynamics, we want to look at an $N$-turn map, given by
\begin{equation}
\mathcal{M}^N = \left ( \mathcal{M}_\perp \mathcal{M}_{s} \right )^N.
\end{equation}
Through a judicious insertion of an identity operator, we can move all of the synchrotron motion maps to the left, and leave only the transverse dynamics to the right. This can be accomplished by noting that
\begin{equation}
\begin{split}
\mathcal{M}_\perp \mathcal{M}_{s} \mathcal{M}_\perp \mathcal{M}_{s} =& \mathcal{M}_\perp \mathcal{M}_{s} \mathcal{M}_{s} \mathcal{M}_{s}^{-1}\mathcal{M}_\perp \mathcal{M}_{s} \\
=& \mathcal{M}_\perp \mathcal{M}_{s}^2 \tilde{\mathcal{M}}_\perp^{(1)}
\end{split}
\end{equation}
where $\tilde{\mathcal{M}}_\perp^{(n)} = \mathcal{M}_s^{-n} \mathcal{M}_\perp \mathcal{M}_s^{n}$. It is straightforward to show that, by moving each successive synchrotron map to the left in this process, we get the $N$-turn map
\begin{equation}\label{eqn:nmap}
\mathcal{M}^N = \left ( \mathcal{M}_s \right )^N \left ( \prod_{n = N}^1 \tilde{\mathcal{M}}_\perp^{(n)} \right )
\end{equation}
where we are counting the index down from left to right.

From the similarity transformation identity described in Appendix~\ref{app:sym_maps}, it is straightforward to compute $\tilde{\mathcal{M}}^{(n)}_\perp$. The similarity transformation moves the synchrotron motion into the argument of the $\mathcal{M}_\perp$ exponential, thus:
\begin{equation}
\begin{split}
\tilde{\mathcal{M}}^{(n)}_\perp =& \mathcal{M}_s^{-n} \mathcal{M}_\perp \mathcal{M}_s^{n} \\
=& \mathcal{M}_s^{-n} \exp \left \{ - \Lieop{\sum_{k=-\infty}^{\infty} h_k(\vec{z}_\perp, A)~e^{i k \psi}} \right \} \mathcal{M}_s^{n} \\
=&  \exp \left \{ - \Lieop{\mathcal{M}_s^{-n} \sum_{k=-\infty}^{\infty} h_k(\vec{z}_\perp, A)~e^{i k \psi}} \right \} \\
=&  \exp \left \{ - \Lieop{ \sum_{k=-\infty}^{\infty} h_k(\vec{z}_\perp, A)~e^{i k \left (\psi - n \mu(A) \right ) }} \right \}
\end{split}
\end{equation}
where $\mu(A)$ is the amplitude-dependent synchrotron phase advance, and we have used the fact that $\mathcal{M}_s \circ (A, \psi) = (A, \psi + \mu(A))$. We now need to compute the product in eqn.~(\ref{eqn:nmap}) as a single exponential operator to first order using the BCH formula from Appendix~\ref{app:sym_maps} to compute the stroboscopic Hamiltonian and its first order correction.

To construct the single exponential operator, we will rely on the BCH formula and a recursion relation defined by concatenating the first $M$ terms from the right of the product in eqn.~(\ref{eqn:nmap}) with the next map to its left. Specifically, let us write the product as
\begin{equation}
 \prod_{n = N}^1 \tilde{\mathcal{M}}_\perp^{(n)} = \left ( \prod_{n = N}^{M+1} \tilde{\mathcal{M}}_\perp^{(n)} \right ) \hat{\mathcal{M}}^{(M)}_\perp
\end{equation}
where
\begin{equation}
\hat{\mathcal{M}}^{(1)}_\perp = \tilde{\mathcal{M}}_\perp^{(1)}
\end{equation}
and
\begin{equation}
\hat{\mathcal{M}}^{(M+1)}_\perp = \tilde{\mathcal{M}}_\perp^{(M)}\hat{\mathcal{M}}^{(M)}_\perp
\end{equation}
so that we end with
\begin{equation}
\hat{\mathcal{M}}^{(N)}_\perp = \prod_{n = N}^1 \tilde{\mathcal{M}}_\perp^{(n)}.
\end{equation}
The goal is therefore to write 
\begin{equation}
\hat{\mathcal{M}}^{(M)}_\perp = \exp \left \{- \lieop{\hat{H}^{(M)}} \right \}
\end{equation}
and compute $\hat{H}^{(M)}$ perturbatively using the BCH series.

From the BCH series, we can derive a recursion relation for $\hat{H}$ to leading order in the Poisson brackets as
\begin{equation}~\label{eqn:h_recur}
\hat{H}^{(M+1)} = \hat{H}^{(M)} + \sum_{k=-\infty}^{\infty} h_k(\vec{z}_\perp, A)~e^{i k \left (\psi - M \mu(A) \right ) } + \varepsilon \frac{1}{2} \left [ \hat{H}^{(M)} , \sum_{k=-\infty}^{\infty} h_k(\vec{z}_\perp, A)~e^{i k \left (\psi - M \mu(A) \right ) } \right ] + \mathcal{O}(\varepsilon^2)
\end{equation}
where we have included $\varepsilon$ to bookkeep the order in Poisson brackets.

Thus to order $\varepsilon = 0$, the stroboscopic Hamiltonian for the $M$-turn map is
\begin{equation}
\begin{split}
\hat{H}^{(M)} =& \sum_{m = 1}^M \sum_{k=-\infty}^{\infty} h_k(\vec{z}_\perp, A)~e^{i k \left (\psi - m \mu(A) \right ) } \\
=&  \sum_{k=-\infty}^{\infty} \sum_{m = 1}^M h_k(\vec{z}_\perp, A)~e^{i k \left (\psi - m \mu(A) \right ) } \\
=&  h_0(\vec{z}, A) \times M + \sum_{k \neq 0}  h_k(\vec{z}_\perp, A)~e^{i k \psi} \left ( \sum_{m = 1}^Me^{- i k m \mu(A)} \right ) \\
=& h_0(\vec{z}, A) \times M + \sum_{k \neq 0}  h_k(\vec{z}_\perp, A)~e^{i k \psi} e^{-i k \mu(A)} \frac{1 - e^{-i k M \mu(A)}}{1 - e^{-i k \mu(A)}} \end{split}
\end{equation}

This sum contains the $k=0$ term, which is the synchrotron phase average $\langle \hat{H}^{(M)} \rangle_\psi = \overline{H}$, while all the other terms are oscillatory in $\psi$ and oscillate with the turn number $M$ with harmonics of the synchrotron frequency. If the synchrotron tune is rational, then the next leading order term is periodic. In this case, terms where $k \mu = 2 \pi K$ will contribute to the sum, the geometric sum will converge to $M$, and we have a stroboscopic invariant to leading order
\begin{equation}
\hat{H}^{(M)} = \left ( h_0 + \sum_{K = \frac{k \mu}{2 \pi}} h_K e^{i K \psi} \right ) \times M
\end{equation}
whereas for irrational tunes $\hat{H}^{(M)}$ is dominated by the $h_0$ term. We define this invariant as
\begin{equation}
\overline{J} = h_0 + \sum_{K = \frac{k \mu}{2 \pi}} h_K e^{i K \psi}.
\end{equation}
If the synchrotron tune is irrational, or the $h_K$ are very small, then $\overline{J}$ is approximately equal to $\overline{H}$, and $\overline{H}$ is an averaged Hamiltonian for the system.


The existence of $\overline{H}$ or $\overline{J}$ suggests the dynamics are well-behaved. Because $h_0$ is independent of $M$, terms with no $\psi$-dependence will vanish from the higher order Poisson brackets, as described in \ref{app:next_order}. Therefore, the higher-order terms in the BCH series contain only terms with $k \neq 0$, and the BCH series for the N-turn map will be of the form
\begin{equation}
\hat{H}^{(N)} = \overline{J} (\vec{z}_\perp, A, \psi) \times N + h(\psi, N)
\end{equation}
where $h$ is oscillatory in $N$ due to the $e^{i k N \mu}$ terms. Furthermore, if the $h_K$ are small or vanishing, then $\overline{J}$ is a pure function of the synchrotron action, which becomes a constant of the motion. On synchrotron tune resonances, there is nevertheless one invariant of the motion, which also constrains the dynamics to isobars of $\overline{J}$. The $N$-turn map is therefore generated by a sum of a secularly growing term, $\overline{J} \times N$, and an oscillatory term, $h$. The dynamics will be dominated by the secular term. 

To see this, we can cast the $N$-turn map as the $N^{th}$ power of a single turn map:
\begin{equation} \label{eqn:strobe_map}
\hat{\mathcal{M}}^{(N)}_\perp \sim \left (\hat{\mathcal{M}}_\perp \right )^N
\end{equation}
with
\begin{equation}
\hat{\mathcal{M}}_\perp = \exp \left ( - \lieop{\overline{H} + \frac{1}{N} h} \right )
\end{equation}
where $h$ is a bounded periodic function of the phase space variables. Thus, in the limit of large $N$, this term becomes perturbatively small compared to $\overline{H}$. This argument, that the oscillatory term does not lead to secular growth in the action, is analogous to the arguments for averaging time-continuous Hamiltonian systems described in \S 19 of Arnold~\cite{arnold_geometrical}.

We remark here on the convergence properties of this series. In the simple limit that $H_\perp$ is uncoupled -- i.e. the chromaticity vanishes -- then the system becomes the uncoupled transverse and longitudinal systems. Furthermore, as $\mu \rightarrow 0$, the value of $\overline{H}$ becomes the exact Hamiltonian. This suggests that the series will converge for sufficiently small synchrotron phase advance. As we show in Appendix~\ref{app:next_order}, the leading order term, and therefore all higher order terms, are oscillatory in $\psi$ and do not grow with turn number. This suggests that the series converges better for large $N$.

\section{Octupole Ring with Chromaticity: A Toy Model}
\label{sec:toy}

To verify our predictions of stroboscopic invariance, we consider a toy ring, comprised of a linear focusing elements with a small octupolar contribution along with a thin linear RF kick. The resulting model produces nonlinear transverse dynamics, including nonlinear chromaticity and synchrotron motion that is analytically tractable.

\subsection{Theoretical Picture}

We model this ring with the product of two symplectic maps:
\begin{equation}
\mathcal{M} = \mathcal{M}_\delta \mathcal{M}_V
\end{equation}
where the Hamiltonians that generate these maps are given by
\begin{equation}
H_\delta = \frac{1}{2} \left ( \mu_x + \xi_1 \delta + \frac{1}{2} \xi_2 \delta^2 \right ) \left ( p^2 + x^2 \right ) + \frac{S_4}{4} x^4 + \frac{1}{2} \alpha_c \delta^2
\end{equation}
and
\begin{equation} \label{eqn:rf_ham}
V(\phi) = \frac{1}{2} V_{rf} \phi^2.
\end{equation}

Breaking this into $\mathcal{M}_\perp$ and $\mathcal{M}_s$, as we do above, gives the two Hamiltonians
\begin{equation}
H_\perp = \frac{1}{2} \left ( \mu_x + \xi_1 \delta + \frac{1}{2} \xi_2 \delta^2 \right ) \left ( p^2 + x^2 \right ) + \frac{S_4}{4} x^4
\end{equation}
and
\begin{equation}
H_s = \frac{1}{2} \mu_s \left ( \hat{\delta}^2 + \hat{\phi}^2 \right )
\end{equation}
where 
\begin{subequations}
	\begin{equation}
		\hat{\delta} = \delta \sqrt{\beta_s} + \frac{\alpha}{\sqrt{\beta_s}} \phi
	\end{equation}
	\begin{equation}
		\hat{\phi} = \frac{\phi}{\sqrt{\beta_s}}
	\end{equation}
\end{subequations}
with the synchrotron Twiss parameters $\beta_s = \frac{\alpha_c}{\sqrt{1 - \frac{1}{2} \alpha_c V_{rf}}}$ and $\alpha_s = \frac{\alpha_c V_{rf}}{2} \frac{1}{\sqrt{1 - \frac{1}{2} \alpha_c V_{rf}}}$, and synchrotron phase advance $\mu_s = \arccos \left ( 1 - \frac{1}{2} \alpha_c V_{rf} \right )$. The synchrotron action is then $A_s = (\hat{\delta}^2 + \hat{\phi}^2)/2$.

Averaging over the synchrotron phase, the stroboscopic Hamiltonian for this system is given by
\begin{equation}
\overline{H} = \frac{1}{2} \left [ \mu_x + \frac{1}{2} \xi_2 \times \left (\frac{1 + \alpha_s^2}{\beta_s} A_s \right ) \right ] \left (p^2 + x^2 \right ) + \frac{S_4}{4} x^4
\end{equation}
We note that the linear chromatic term does not contribute in this example, because for linear synchrotron motion $\langle \delta \rangle_\theta = 0$. This is specific to linear synchrotron motion, and does not hold with nonlinear rf effects. To demonstrate that this is the correct stroboscopic Hamiltonian, we will compare particle trajectories from integrating the full model to trajectories computed using the stroboscopic Hamiltonian -- as noted in eqn.~(\ref{eqn:strobe_map}) the stroboscopic Hamiltonian generates the transverse dynamics with a period of a synchrotron oscillation.

\subsection{Numerical Model}

To generate the numerical data, we have to integrate the individual maps. The synchrotron map and the transverse map, in the limit of $S_4 \rightarrow 0$, can be solved analytically. To include the octupole term, we use a fourth order splitting of $\mathcal{M}_\perp$ using the identity described by Yoshida~\cite{yoshida:90}. As a benchmark, we look at variations in $H_\perp$ for an on-momentum particle, which is exactly conserved given an exact solution to $H_\perp$.

\begin{figure}[!htp]
\centering
  \includegraphics[width=0.7\columnwidth]{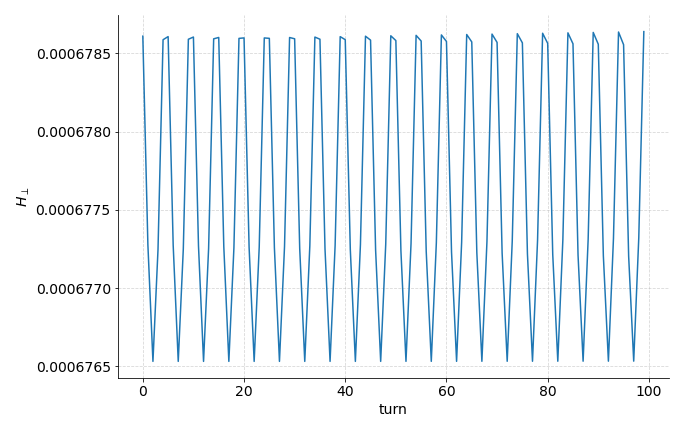}
  \caption{On-momentum $H_\perp$ invariant calculated using the fourth order scheme for a particle with no longitudinal dynamics.}
  \label{fig:benchmark}
\end{figure}

The operator splitting requires a second-order integration, which splits $H_\perp$ into the transverse linear component, and the octupole kick. Using the Yoshida identity transforms this to a fourth order operation, reducing the numerically induced error in $H_\perp$ conservation. As we can see in fig.~(\ref{fig:benchmark}), $H_\perp$ is conserved on the order of .2\%. This provides a baseline for our assessment, and any variation in $H_\perp$ varying above this level should indicate a physical outcome rather than a numerical artifact of our integration scheme.

\subsection{Results}

For the following simulations, we use parameters designed to emulate the tunes of the iRCS lattice described in Section~\ref{sec:ircs} below. In particular, we use the numerical values in Table~(\ref{tab:toy_params}). For these parameters, the lattices have the same linear betatron tune, synchrotron tune, and linear chromaticity.

\begin{table}[!htp]
\centering
\begin{tabular}{| l | c |}
\hline
Parameter & Value \\
\hline
\hline
$\mu_x$ & $21.6 \times 2\pi$ \\
$\xi_1$  & -79. \\
$\xi_2$ & $140.8$ \\
$S_4$ & $100.1$ \\
$V_{rf}$ & .42 \\
$\alpha_c$ & $.59$ \\
\hline
$\beta_s$ & $1.2 \times 10^{-3}$ \\
$\alpha_s$ & $.257$ \\
$\nu_s$ & .08 \\
\hline
\end{tabular}
\caption{Parameters of the toy octupole ring with synchrotron motion.}
\label{tab:toy_params}
\end{table}

\begin{figure}[ht]
  \centering
  \includegraphics[width=0.75\columnwidth]{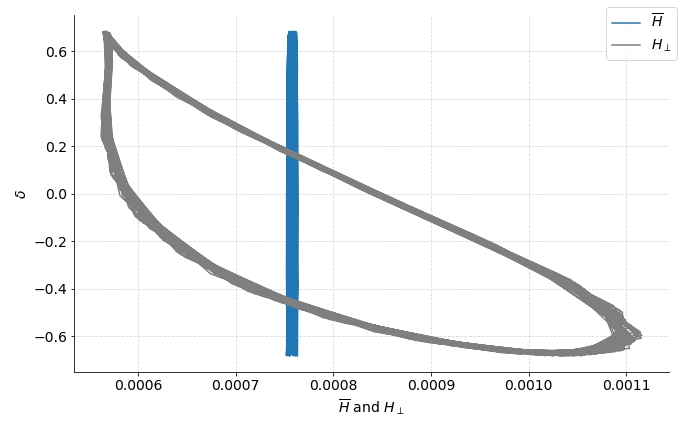}  
\caption{Comparison of the coasting beam Hamiltonian $H_\perp$ and stroboscopic Hamiltonian $\overline{H}$ as the momentum varies over 500 turns, or approximately 40 synchrotron oscillations.}
\label{fig:hams}
\end{figure}

The resulting evolution in $H_\perp$ and $\overline{H}$ are shown as a function of $\delta$ in figure~(\ref{fig:hams}). The near-closed-loop periodicity in $H_\perp$ with $\delta$ suggests the existence of some invariant of the motion. $\overline{H}$ shows much smaller variation with $\delta$, suggesting that it is an approximation to that invariant -- the scale of variation of $H_\perp$ is an order of magnitude larger than the scale of variation of $\overline{H}$. It is furthermore worth noting that $A_s$ is not a turn-by-turn invariant due to the chromatic coupling, so all of the quantities used to compute $\overline{H}$ are varying turn-by-turn.

\begin{figure}
  \centering
  \includegraphics[width=0.75\columnwidth]{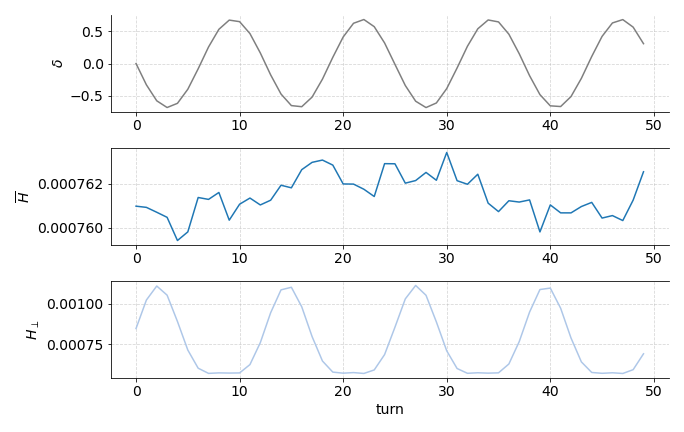}  
  \caption{Turn-by-turn comparison of $H_\perp$ and $\overline{H}$ with linear synchrotron motion.}
  \label{fig:comped_ham}
\end{figure}

Comparing $\overline{H}$ to $H_\perp$ with the synchrotron motion, as we do in figure~(\ref{fig:comped_ham}), we can see that the stroboscopic Hamiltonian is very well-conserved throughout the synchrotron oscillation, while $H_\perp$, which generates the turn-by-turn variation, varies periodically with the synchrotron motion. This periodicity with the synchrotron motion is suggestive of underlying integrable motion —- if a system is integrable, any dynamical quantity will exhibit some quasi-periodicity with the dynamics~\cite{lichtenbergLieberman:text}. In our case, $H_\perp$ is periodic with the synchrotron motion, which suggests the existence of an underlying integrable Hamiltonian which has as one set of action-angle variables the synchrotron action and phase. That $\overline{H}$ is extremely well-conserved suggests that $\overline{H}$ is that Hamiltonian.

To confirm this, we compare the Poincar\'e sections of trajectories generated by $\overline{H}$ and compare to the full coupled dynamics to see if they trace out the same trajectory in phase space. We expect differences, as in this case the turn-by-turn variation in the betatron phase advance due to the chromaticity is replaced with a constant betatron tune shifted by the second-order chromatic term. However, if $\overline{H}$ approximates the long term dynamics with an oscillatory correction, we would expect the trajectories of $\overline{H}$ to track the smooth motion of the full dynamics.

\begin{figure}
  \centering
  \includegraphics[width=0.75\columnwidth]{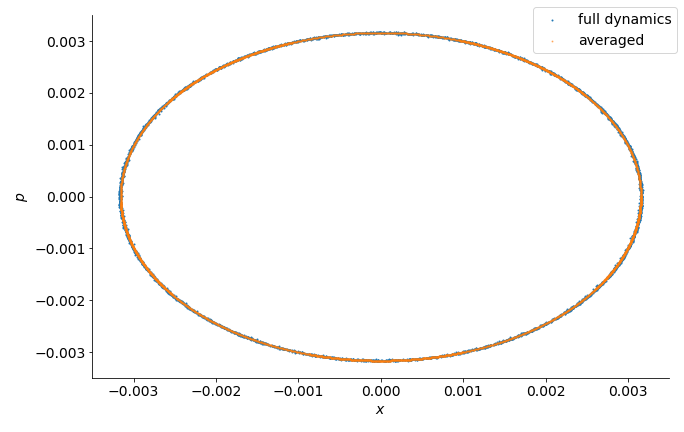}  
  \caption{Poincar\'e section for the full dynamics (blue) and the dynamics of the averaged Hamiltonian (orange) over 5000 turns.}
  \label{fig:hams_comp}
\end{figure}

We can see this in fig.~(\ref{fig:hams_comp}), which compares the transverse tracking data from integrating the full system (blue) or the dynamics of $\overline{H}$ (orange) over 5000 turns. What we see is that they both track out basically the same ellipse in phase space, with the blue trajectories showing some small-scale variability compared to the averaged dynamics. This suggests that the transverse dynamics of $\overline{H}$ captures a smoothed version of the full dynamics.

This picture is unchanged with nonlinear rf effects, such as amplitude-dependent synchrotron tune. In this example, we will look at the dynamical quantities computed using the linear rf dynamics, so we are looking at an approximate averaged Hamiltonian that lacks higher order terms due to the rf curvature.

\begin{figure}
  \centering
  \includegraphics[width=0.75\columnwidth]{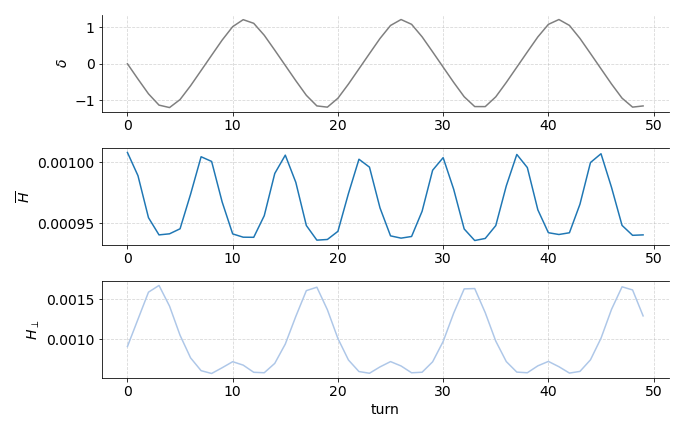}  
  \caption{Turn-by-turn comparison of $H_\perp$ and $\overline{H}$ with nonlinear synchrotron motion.}
  \label{fig:nl_rf}
\end{figure}

In figure~(\ref{fig:nl_rf}), we see the particle motion at a larger initial $\phi$ and with the rf potential $\frac{1}{2} V_{rf} \phi^2$ replaced with $V_{rf} \cos \phi$ in eqn.~(\ref{eqn:rf_ham}). We can see a stronger second harmonic oscillation in both $H_\perp$ and $\overline{H}$, but as with the case in figure~(\ref{fig:comped_ham}), $\overline{H}$ is very nearly conserved, with variations at the $\sim1\%$ level, while $H_\perp$ has oscillations at the $\sim25-50\%$ level.

We expect this, as the chromatic terms in $H_\perp$ do not contain the corrections to the linear rf action-angle variables that come from including rf curvature. We see a clear periodicity in $\overline{H}$, suggesting that a higher order perturbative correction that cancels those oscillations in $A$ would give better conservation to $\overline{H}$.

\section{Stroboscopic Invariants in an Integrable Rapid-Cycling Synchrotron}
\label{sec:ircs}

To observe the existence of stroboscopic invariants in a complex system, we consider an integrable rapid-cycling synchrotron~\cite{eldred_valishev:2017} (iRCS). This lattice design includes nonlinear integrable dynamics~\cite{danilov_nagaitsev:10,nag_val_dan_10}, which intrinsically includes a tune spread with transverse amplitude designed to Landau damp coherent instabilities. Because our prior analysis is independent of the nature of the transverse Hamiltonian dynamics in the lattice, we expect to see two stroboscopic invariants of the motion. 

The iRCS transverse dynamics with synchrotron motion is quite complicated. As described by Webb et al.~\cite{webb_etal:15c}, we expect to see integrable behavior for momenta in ranges where the vertical and horizontal chromaticity are equal. In these regions we would therefore expect the beam dynamics to be much better behaved in general. Computing the stroboscopic Hamiltonian is difficult for the full lattice, as it must include a nonlinear normal form analysis of the transverse dynamics with momentum spread as well as the longitudinal dynamics.

We must therefore look for properties we would expect from the existence of an underlying constant of the motion, such as periodicity in the behavior of dynamical variables commensurate with the periodicity we consider for the stroboscopic Hamiltonian, i.e. the synchrotron period.

To test this periodicity prediction, we computed the on-momentum ($A = 0$) Danilov-Nagaitsev invariants from~\cite{danilov_nagaitsev:10} through many synchrotron oscillations. For small-amplitude synchrotron oscillations, we expect the effects of finite $A$ to be perturbative, and we will see a synchrotron motion periodicity with the Danilov-Nagaitsev invariants.

Table~\ref{tab:iRCS_params} shows the key parameters for this lattice design. The phase advance through the nonlinear insert of $Q_{0}=0.3$, the nonlinear strength parameter is {$t=0.3$, and elliptic potential parameter is $c=0.14$ m$^{1/2}$.} (see \cite{danilov_nagaitsev:10,nag_val_dan_10}).

The iRCS is designed with 1.680 MV total RF voltage to provide to achieve a 20 Hz ramp rate and a 8 GeV extraction energy. In application, every other cell of the iRCS would contain RF cavities and the harmonic number for the ring would be 113. For modeling purposes, each of the twelve periodic cells has RF cavities providing 140 kV and the harmonic number for the ring is $9\times12=108$. To avoid transition crossing, the momentum compaction factor of the iRCS is designed to be $5.9 \times 10^{-4}$. At the injection energy 0.8 GeV, the synchrotron tune for the ring is 0.08 (and 0.007 per periodic cell).

\begin{figure}[!htp]
\centering
  \includegraphics[width=0.7\columnwidth]{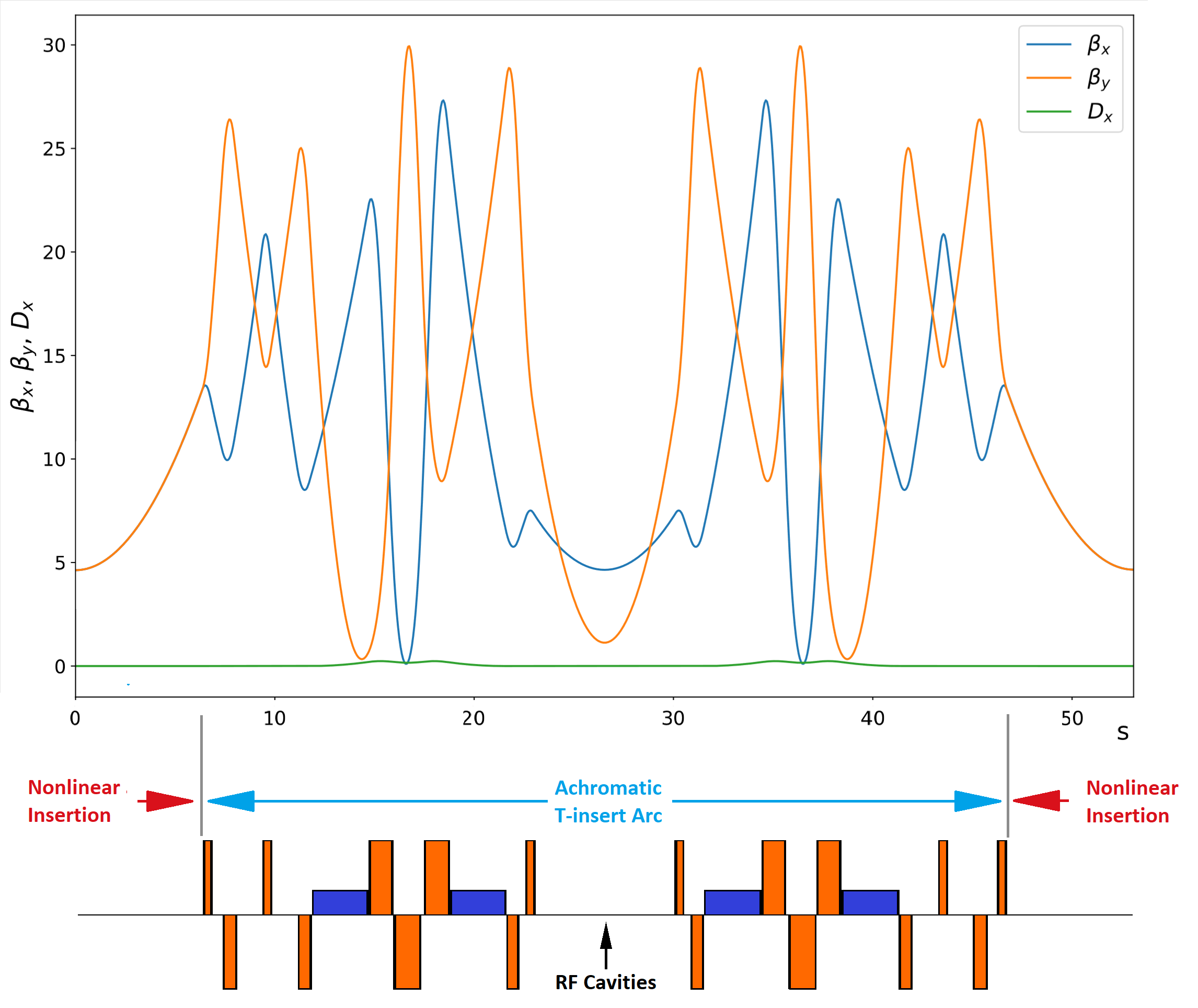}
  \caption{(top) Twiss parameters for one of the twelve periodic cells. (bottom) Beamline layout where dipoles are shown as short blue rectangles and quadrupoles as tall orange rectangles.}
  \label{fig:iRCS_lattice}
\end{figure}


\begin{table}[!htp]
\centering
\caption{Parameters of iRCSv3 Lattice}
\begin{tabular}{| l | c |}
\hline
Parameter & Value \\
\hline
\hline
Circumference & 636 m \\
Periodicity & 12 \\
Bend Radius & 15.4 m \\
Max Beta Function  & 30 m \\
Max Dispersion & 0.22 m \\
\hline
Betatron Tune & 21.6 \\
Linear Chromaticity & -79 \\
Momentum Compaction & 5.9 $\times$ $10^{-4}$ \\
Insertion lengths per cell & 7.2 m, 4 $\times$ 1.3m \\
RF Voltage & 1.680 MV \\
Synchrotron Tune & 0.08 \\
\hline
NL Insertion Length & 12.7 m \\
Phase-advance over insert & $0.3 \times \pi$ \\
Nonlinear Strength t-value & 0.3 \\
Elliptic Distance c-value & 0.14 m$^{\frac{1}{2}}$ \\
\hline
95\% Transverse Emittance & 20 mm mrad \\
95\% Longitudinal Emittance & 0.09 eV$\cdot $s \\
Vertical Lattice Tune Spread & 0.52 \\
Horizontal Lattice Tune Spread & 0.34 \\
Chromatic Tune spread & 0.52 \\
\hline
\end{tabular}
\label{tab:iRCS_params}
\end{table}

\begin{figure}[htp]
\centering
  \includegraphics[width=0.7\columnwidth]{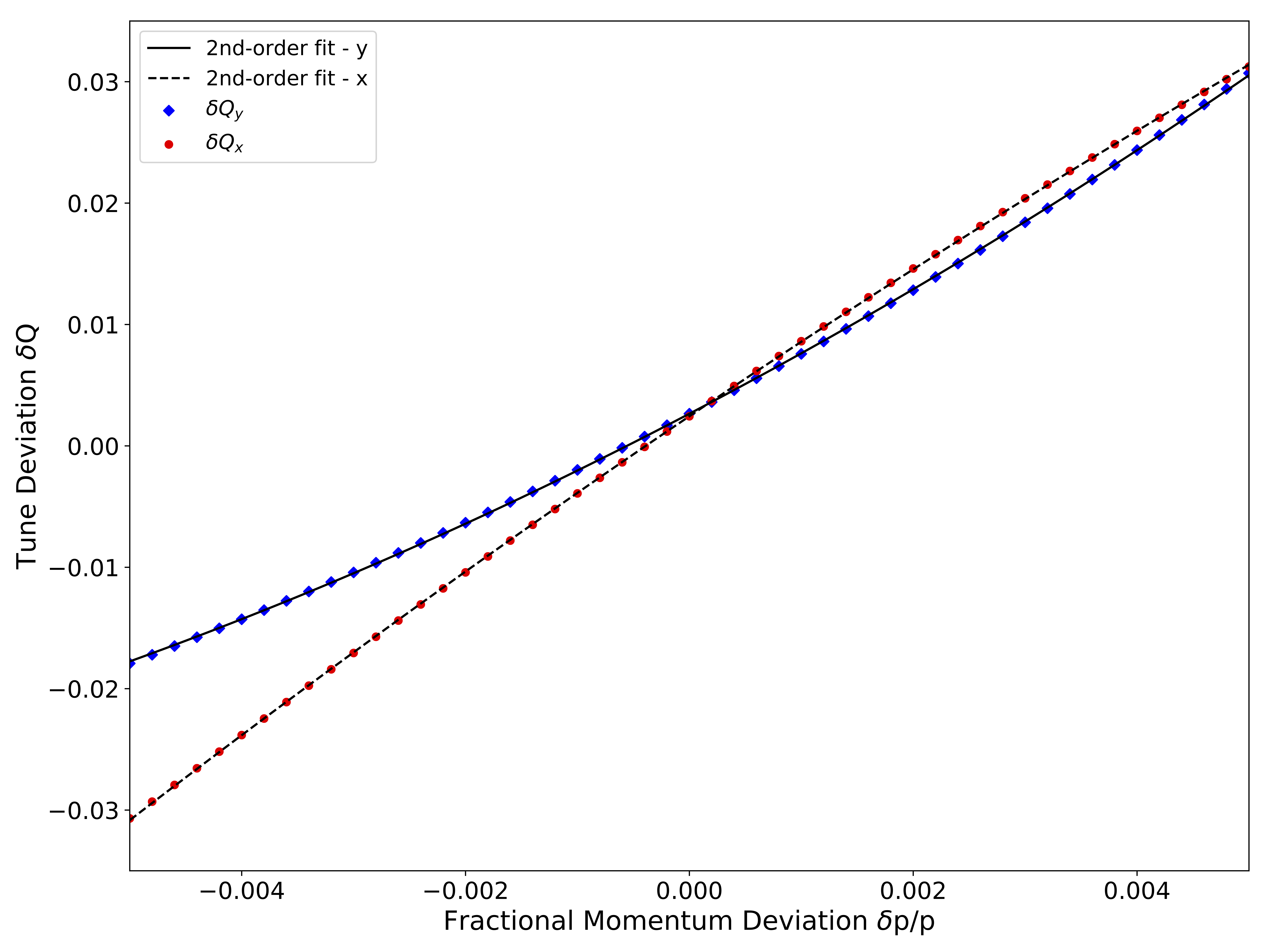}
  \caption{Vertical and horizontal chromaticities are plotted for a single cell of the 12-cell iRCS design.}
  \label{fig:chromaticity}
\end{figure}

The iRCS lattice was optimized to control the discrepancy between the horizontal and vertical tune across the momentum span $\pm 0.5 \%$ without the use of sextupoles. The iRCS lattice also has the flexibility to finely adjust the betatron tune-matching and chromaticity matching independently. Figure~\ref{fig:chromaticity} shows the tune dependence on momentum, measured by tracking the small-amplitude betatron oscillation of off-momentum particles, with the strength of the elliptic element set to zero.

The chromaticity combined with the nonlinear integrable optics makes the iRCS lattice a fairly complex example of synchro-betatron coupling.

\begin{figure}[htbp]
\centering 
\includegraphics[width=.75\textwidth]{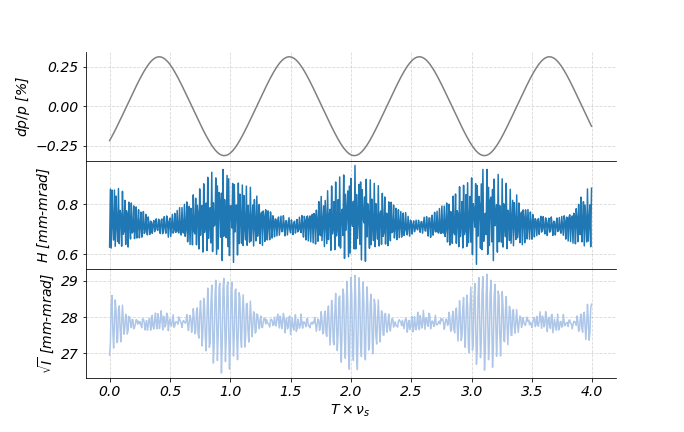}
\caption{Sample trajectory showing oscillations in the on-momentum invariants in an integrable RCS.}\label{fig:stack}
\end{figure}

In fig.~(\ref{fig:stack}) we plot the particle momentum offset and on-momentum Danilov-Nagaitsev invariants against the turn number $T$ times the zero-amplitude synchrotron tune $\nu_s$. As we can see, there is oscillatory behavior in the invariants periodic with the synchrotron oscillation, indicating the existence of a stroboscopic invariant. We can also see that this is in a regime where there is a finite amplitude-dependent synchrotron tune depression, as the successive minima in the top plot of fig.~(\ref{fig:stack}) are slightly greater than $T \times \nu_s = 1$ separated, indicating that $\nu_s(A) > \nu_s(0)$.

This periodicity is consistent over many hundreds of synchrotron periods, and across many initial particle trajectories. That persistence without secular growth indicates the presence of stroboscopic invariants in the Danilov-Nagaitsev Hamiltonian with synchrotron motion and chromatic effects, as applied to the iRCS lattice.

\section{Discussion}
\label{sec:discussion}

We have presented an approach to computing the $N$-turn map for an entire synchrotron period, and derived a stroboscopic Hamiltonian $\overline{H}$ which defines the secular Hamiltonian dynamics of the full synchro-betatron coupling. The stroboscopic Hamiltonian is the average of the transverse Hamiltonian over a synchrotron period. This Hamiltonian is $\mathcal{O}(\nu_s^{-1})$, with $\nu_s \ll 1$ the synchrotron tune, while the correction terms remain $\mathcal{O}(1)$. Therefore, this holds well for small synchrotron tune. Furthermore, for multi-synchrotron-period maps, these correction terms will oscillate with the number of periods, while the stroboscopic Hamiltonian term will grow linearly, suggesting that it dominates the long-term dynamics. We presented evidence of this stroboscopic Hamiltonian in the context of an integrable optics rapid cycling synchrotron, showing that the on-momentum Danilov-Nagaitsev invariants vary with momentum offset, but are periodic with the synchrotron period. The result, however, is generic to any Hamiltonian for the transverse dynamics, so long as a single Hamiltonian which generates the single-turn map for the transverse dynamics exists, i.e. in the absence of chaos.

\appendix
\section{Symplectic Maps, Lie Algebras, and the Baker-Campbell-Hausdorff Formula}
\label{app:sym_maps}

In this Appendix we will overview the mathematics of symplectic maps and Lie operators, highlighting key mathematical identities that we will use in this paper. Much of this is a survey of prior work by Dragt and others~\cite{dragt_abell:96,dragt_finn:1976,dragt_forest:83,dragt_text,dragt:79,dragt:82,dragt:84} as it pertains to the work presented here. We omit proofs for the sake of brevity, opting to state the relevant identities.

Given a Hamiltonian $H$, the equations of motion for a particle's phase space trajectory will satisfy the Poisson bracket differential equation
\begin{equation}
\frac{d \vec{z}}{dt} = - [H, \vec{z}]
\end{equation}
We can interpret $[H, \star]$ as a \emph{Lie operator} that acts on $\vec{z}$, denoted by $\lieop{H}$. This implies that the evolution of $\vec{z}$ can be cast as an operator differential equation, with the flow $\vec{z}_f = \mathcal{M}_{i \rightarrow f} \vec{z}_i$. This leads to the operator differential equation for the symplectic map $\mathcal{M}$ which describes the flow for the Hamiltonian $H$:
\begin{equation}
\frac{d}{dt} \mathcal{M}_{t_i \rightarrow t} = - \lieop{H} \mathcal{M}_{t_i \rightarrow t}
\end{equation}
with the initial condition $\mathcal{M}_{t_i \rightarrow t_i} = \mathcal{I}$, the identity. $\mathcal{M}$ contains all of the dynamics for the Hamiltonian $H$. We can solve this operator equation by iterative integration, i.e.\begin{equation}
\mathcal{M}_{t_i \rightarrow t} = \mathcal{I} - \int_{t_i}^t dt' ~ \lieop{H} \mathcal{M}_{t_i \rightarrow t'}
\end{equation}
Assuming $H$ is independent of time, the solution can be written as the exponential operator
\begin{equation}
\begin{split}
\mathcal{M}_{t_i \rightarrow t} =& \sum_{n = 0}^{\infty} \frac{(-1)^n}{n!} (t - t_i)^n \lieop{H}^n \\
\equiv& \exp \left \{ - \lieop{H}(t - t_i) \right \}
\end{split}
\end{equation}
where $\lieop{H}^n$ is defined as repeated application of the operator $\lieop{H}$. Operator exponentials play an important role in Lie algebraic treatments of symplectic maps.

In a particle accelerator, a symplectic map describes the change of phase space coordinates at the exit of the element given the coordinates at the entrance of the element:
\begin{equation}
\vec{z}_{\textsf{out}} = \mathcal{M}_i \circ \vec{z}_{\textsf{in}}
\end{equation}
In a ring, the product of all of these symplectic maps forms the \emph{single-turn map}
\begin{equation}
\mathcal{M} = \prod_i \mathcal{M}_i
\end{equation}
which contains the full dynamics of the ring. Computing this single-turn map is the subject of normal form analysis and Taylor and Cremona polynomials,  as well as, indirectly, the goal of tracking codes. For the purposes of this paper, we assume that we have already calculated the single-turn map, and that it is of the form
\begin{equation}
\mathcal{M} = e^{-\lieop{H}}
\end{equation}
where $-H$ is the generator of the map. This Hamiltonian is related to the invariants of motion, such as the Courant-Snyder invariants or the Danilov-Nagaitsev Hamiltonian.

The computation in this paper relies on two identities for these maps: the similarity transformation, and the Baker-Campbell-Hausdorff formula.

The similarity transform states that
\begin{equation}
\mathcal{G}^{-1} \lieop{f} \mathcal{G} = \lieop{\mathcal{G} f}.
\end{equation}
This identity frequently appears in the context of coordinate transformations, but in our case arises as we move all the synchrotron motion maps to the left. It is straightforward to show that
\begin{equation}
\mathcal{G}^{-1} \lieop{f}^n \mathcal{G} = \lieop{\mathcal{G} f}^n.
\end{equation}
by judicious insertion of $\mathcal{G} \mathcal{G}^{-1}$ between each instance of $\lieop{f}$, and we can therefore see that
\begin{equation}
\mathcal{G}^{-1} e^{\lieop{f}} \mathcal{G} =  e^{\lieop{\mathcal{G} f}}
\end{equation}

The Baker-Campbell-Hausdorff (BCH) formula provides a procedure for combining two non-commuting exponential Lie maps into one exponential Lie map, by defining a series for the generator of that combined Lie map. If we want to write the product of two exponential Lie maps as a single exponential Lie map, $e^{\lieop{f}} e^{\lieop{g}} = e^{\lieop{h}}$, then the BCH formula tells us the series for $h$ in terms of $f$ and $g$:
\begin{equation}{\label{eqn:bch}}
h = f + g + \frac{1}{2} [f, g] + \frac{1}{12} \left ( [f, [f, g]] - [g, [f, g]] \right ) + \dots
\end{equation}
Although this is a formal power series, it may be asymptotic and indeed may not converge at all. We therefore need $[f, g]$ to be in some sense ``small''. This can mean multiple things, and the BCH series can be a perturbation series in, for example, powers of $\vec{z}$ in the multipole picture of particle accelerators, or in this case the synchrotron tune, as we discuss in Appendix~\ref{app:next_order}.

\section{Leading Order Correction}
\label{app:next_order}

To compute the next-leading order term for finite synchrotron tune, we need to go to the next order in the BCH series. We will truncate the series at $\mathcal{O}(\varepsilon)$, so that we only consider single pairwise Poisson brackets. From eqn.~(\ref{eqn:h_recur}), we can add a term so that we are computing $\hat{H}^{(M)} + \varepsilon P^{(M)}$, where $P$ is the next order Poisson bracket term. This immediately gives the recursion relation:
\begin{equation}
\varepsilon P^{(M+1)} = \varepsilon P^{(M)} + \frac{1}{2} \varepsilon \left [ \hat{H}^{(M)}, \sum_k h_k e^{i k (\psi - M \mu)} \right ]
\end{equation}
with the initial condition that $P^{(1)} = 0$. Therefore, we have that
\begin{equation}
P^{(N)} = \frac{1}{2} \sum_{n = 1}^N \left [ \hat{H}^{(n)}, \sum_k h_k e^{i k (\psi - n \mu)} \right ]
\end{equation}
and furthermore, from our solution of the leading order Hamiltonian $\hat{H}^{(n)}$, we have
\begin{equation}
P^{(N)} = \frac{1}{2} \sum_{n = 1}^N \sum_{m = 1}^n \sum_{k, k'} \left [ h_{k'} e^{i k' \psi}e^{- i k' m \mu}, h_k e^{i k \psi} e^{- i k n \mu)} \right ].
\end{equation}
For clarity, define $f_k = h_k e^{i k \psi}$ and get that
\begin{equation}
P^{(N)} = \frac{1}{2} \sum_{n = 1}^N \sum_{m = 1}^n \sum_{k, k'} \left [ f_{k'} e^{- i k' m \mu}, f_k e^{- i k n \mu} \right ]
\end{equation}
The $(k, k') = 0$ term vanishes, so the only surviving terms in this correction oscillate in harmonics of the synchrotron period, due to the $e^{i q \mu}$-type terms in the series. This means that this perturbation remains bounded as compared to the secular growth of the averaged Hamiltonian term $\overline{H}$.

\acknowledgments

This work was supported in part by the United States Department of Energy, Office of Science, Office of High Energy Physics under contract no. DE-SC0011340 and in part by Fermi Research Alliance, LLC under Contract No. DE-AC02-07CH11359 with the United States Department of Energy.


\bibliography{stroboscopic_hamiltonian.bib}

\end{document}